\documentclass[twocolumn]{aa}

\usepackage{graphicx}
\usepackage[varg]{txfonts}
\usepackage{amsmath}

\newcommand{\tb}{$\tau$ Boo}

\newcommand{\smallvspace}{\vspace{0.1cm}}

\newcommand{\nwalkers}{96}
\newcommand{\nsteps}{$3.6 \cdot 10^6$}
\newcommand{\totalsamples}{$346 \cdot 10^6$}
\newcommand{\nburn}{$1.2\cdot 10^6$}
\newcommand{\nthin}{$6$}

\newcommand{\specA}{F6V}
\newcommand{\VA}{$4.5$}
\newcommand{\RAA}{13:47:15.737}
\newcommand{\DECA}{+17:27:24.79}
\newcommand{\parallax}{$63.9\pm 0.3$}
\newcommand{\MA}{$1.35\pm 0.03$}
\newcommand{\RA}{$1.42\pm 0.02$}
\newcommand{\TeffA}{$6387\pm 44$}
\newcommand{\FeHA}{$0.25\pm 0.03$}
\newcommand{\vsiniA}{$15.0 \pm 0.5$}
\newcommand{\loggA}{$4.26\pm0.06$}
\newcommand{\ageA}{$1.3^{+0.4}_{-0.6}$}
\newcommand{\ProtA}{$3.2 \pm 0.2$}

\newcommand{\inclAb}{$44.5 \pm 1.5$\,deg}

\newcommand{\logaarc}{$1.149^{+0.21}_{-0.14}$}
\newcommand{\tperiAD}{$2026.9^{+1.4}_{-1.7}$}
\newcommand{\tperi}{\tperiAD}
\newcommand{\tperiMJD}{$61366^{+529}_{-604}$}
\newcommand{\ecosw}{$0.330^{+0.182}_{-0.157}$}
\newcommand{\esinw}{$-0.873^{+0.114}_{-0.067}$}
\newcommand{\bigomega}{$191.82^{+ 3.3}_{-4.7}$}
\newcommand{\cosi}{$0.680^{+0.046}_{-0.036}$}
\newcommand{\MB}{$0.49\pm 0.02$}
\newcommand{\gaa}{$-647.4^{+219.0}_{-181.1}$}
\newcommand{\gab}{$43.9^{+10.2}_{-10.0}$}
\newcommand{\gac}{$-12.6^{+7.2}_{-7.0}$}
\newcommand{\gad}{$-16179.3^{+17.3}_{-17.0}$}
\newcommand{\Kp}{$468.42\pm 2.09$}
\newcommand{\Pp}{$3.312454\pm 3.3\cdot 10^{-6}$}
\newcommand{\tp}{$56401.8797 \pm 0.0036$}
\newcommand{\distance}{$15.66\pm 0.08$}
\newcommand{\jrhoold}{$0.245^{+0.027}_{-0.022}$}
\newcommand{\jrhonew}{$0.054^{+0.023}_{-0.015}$}
\newcommand{\jthetaold}{$1.24^{+0.13}_{-0.12}$}
\newcommand{\jthetanew}{$0.634^{+0.50}_{-0.30}$}
\newcommand{\jrva}{$17.6^{+2.9}_{-2.5}$}
\newcommand{\jrvb}{$15.3^{+3.4}_{-2.8}$}
\newcommand{\jrvc}{$26.5^{+2.5}_{-2.2}$}
\newcommand{\jrvd}{$12.2^{+2.2}_{-1.7}$}

\newcommand{\incl}{$47.2^{+2.7}_{-3.7}$}
\newcommand{\aAU}{$221^{+138}_{-62}$}
\newcommand{\aAUperi}{$28.3^{+2.3}_{-3.0}$} 
\newcommand{\Pyr}{$2420^{+2587}_{-947}$}
\newcommand{\ecc}{$0.87^{+0.04}_{-0.03}$}
\newcommand{\athreePtwo}{$1.84\pm 0.03$}
\newcommand{\smallomega}{$290.7^{+13}_{-10}$}

\newcommand{\RB}{$0.48\pm 0.05$}
\newcommand{\TeffB}{$3580 \pm 90$}

\begin{document} 
\title{Constraining the orbit of the planet-hosting binary $\tau$ Boötis}
\subtitle{Clues about planetary formation and migration}
\author{A.\ B.\ Justesen\inst{1}\fnmsep\thanks{\email{justesen@phys.au.dk}}
        \and S.\ Albrecht\inst{1}}

\institute{Stellar Astrophysics Centre, Department of Physics and Astronomy, Aarhus University, Ny Munkegade 120, DK-8000 Aarhus C, Denmark.}

\date{Received \today; accepted \today}

\abstract
{The formation of planets in compact or highly eccentric binaries and the migration of hot Jupiters are two outstanding problems in planet formation. Detailed characterisation of known systems is important for informing and testing models. The hot Jupiter \tb\ Ab orbits the primary star in the long-period ($P \gtrsim 1000$\,\textrm{yr}), highly eccentric ($e \sim 0.9$) stellar double star system $\tau$ Boötis. Due to the long orbital period, the orbit of the stellar binary is poorly constrained.}
{Here we aim to constrain the orbit of the stellar binary \object{\tb}\ AB in order to investigate the formation and migration history of the system. The mutual orbital inclination of the stellar companion and the hot Jupiter has important implications for planet migration. The binary eccentricity and periastron distance are important for understanding the conditions under which \object{\tb\ Ab} formed.}
{We combine more than 150 years of astrometric data with twenty-five years of high precision radial velocities. The combination of sky-projected and line-of-sight measurements places tight constraints on the orbital inclination, eccentricity, and periastron distance of \tb\ AB.}
{We determine the orbit of \tb\ B and find an orbital inclination of \incl\,deg, periastron distance of \aAUperi\,\textrm{au} and eccentricity of \ecc. We find that the orbital inclinations of \tb\ Ab and \object{\tb\ B}, as well as the stellar spin-axis of \object{\tb\ A} coincide at $\sim \! \! \, 45$ degrees, a result consistent with the assumption of a well-aligned, coplanar system.}
{The likely aligned, coplanar configuration suggests planetary migration within a well-aligned protoplanetary disc. Due to the high eccentricity and small periastron distance of \tb\ B, the protoplanetary disc was tidally truncated at $\approx \! 6$\,\textrm{au}. We suggest that \tb\ Ab formed near the edge of the truncated disc and migrated inwards with high eccentricity due to spiral waves generated by the stellar companion.}


\maketitle

\section{Introduction} \label{sec:intro}
The origin of hot Jupiters is one of the longest-standing puzzles in exoplanetary science. The three most likely formation mechanisms -- disc migration, high eccentricity migration and \textit{in situ} formation -- all have theoretical and observational shortcomings (see \citet{2018arXiv180106117D} for a recent review). High-$e$ migration theories propose migration triggered by dynamical interactions such as planet-planet scattering or Kozai-Lidov oscillations with subsequent tidal circulation \citep[e.g.][]{1996Natur.384..619W, 2003ApJ...589..605W}. During these interactions, the orbital eccentricity and inclination are excited to large values. High-$e$ migration is motivated by measurements showing that a significant fraction of hot Jupiters orbiting F-type stars are misaligned with the spin-axis of their host stars \citep{2010ApJ...718L.145W, 2012ApJ...757...18A}. For hot Jupiters orbiting in binary (or multiple) stars, the stellar companion is a possible source of such interactions. One might expect that planets in compact or highly eccentric binaries (periastron distance $a_{\rm peri} < 30$\,\textrm{au}) are likely candidates for having experienced this type of migration. However, the number of planets in such systems are expected to be limited due to star-disc interactions. The presence of a close stellar companion affects the protoplanetary disc, truncating the disc at a fraction of the periastron distance \citep{1994ApJ...421..651A}. This effect reduces the disc mass and increases impact velocities within the disc, possibly inhibiting or preventing planet formation. Furthermore, discs in compact binaries show significantly reduced disc luminosities (\textit{i.e.} masses) compared to single stars \citep{2012ApJ...751..115H}. For these reasons, it has been suggested that giant planets are unlikely to form in small-separation binaries \citep{1996ApJ...458..312J, 2012ApJ...751..115H}. Nevertheless, the discovery of several planetary systems in compact binaries have shown that planets do form in such systems, contrary to expectation (e.g. $\gamma$ Cephei A \citep{2003ApJ...599.1383H} and Kepler-444A \citep{2016ApJ...817...80D}). To understand the formation of planets in compact binaries, precise orbital parameters of both the planetary and stellar orbits are needed. In particular, the inclination, eccentricity, periastron distance and mass ratio of the binary all impact the evolution of the disc as well as the formation and migration of planets in the system.

As a hot Jupiter orbiting an F-type star in an eccentric binary, \tb\ Ab is an ideal target for testing both hot Jupiter migration and planet formation in binaries. \tb\ is a bright ($V=$ \VA) planet-hosting double star. The binary contains the hot, massive star \tb\ A (F6V) orbited by the smaller companion \tb\ B (M2). \tb\ A hosts the hot Jupiter \tb\ Ab in a 3.3\,day orbit \citep{1997ApJ...474L.115B}. The stellar companion \tb\ B is in an eccentric, long period orbit. The most recent estimates of the binary orbit are based on astrometric data and suggest a highly eccentric orbit $e \sim 0.8$ with a period of $\sim 1000$\,\textrm{yr} \citep{2011AJ....142..175R, 2014AJ....147...65D}. Due to the long period, astrometric measurements cover less than $20\%$ of orbital period. Both Kozai-Lidov oscillations \citep{2014IAUS..299...52M} and migration within an eccentric disc \citep{2011AJ....142..175R} have been suggested as possible migration mechanisms in the \tb\ system. However, the poorly constrained binary orbit has prevented any conclusive determination of the formation and migration scenario.

If the \tb\ system interacted dynamically and experienced high-$e$ migration triggered by Kozai-Lidov oscillations, the planet and stellar companion are expected to orbit on mutually inclined orbits. If the planet and companion star migrated within an aligned disc, the system is expected to be coplanar. The orbital alignment between \tb\ Ab and \tb\ B is therefore crucial for understanding the formation of the system. Similarly, the eccentricity and periastron distance of \tb\ B determine the size of the protoplanetary disc in which \tb\ Ab formed. 

We investigate the possible formation and migration mechanisms of the \tb\ system by precisely characterising the orbit of the \tb\ AB binary. This is achieved by combining more than 150 years of astrometric data with 20 years of high precision radial velocities.

\section{The \tb\ System} \label{sec:system}

\begin{figure*}
    \centering
    \includegraphics[width=\textwidth]{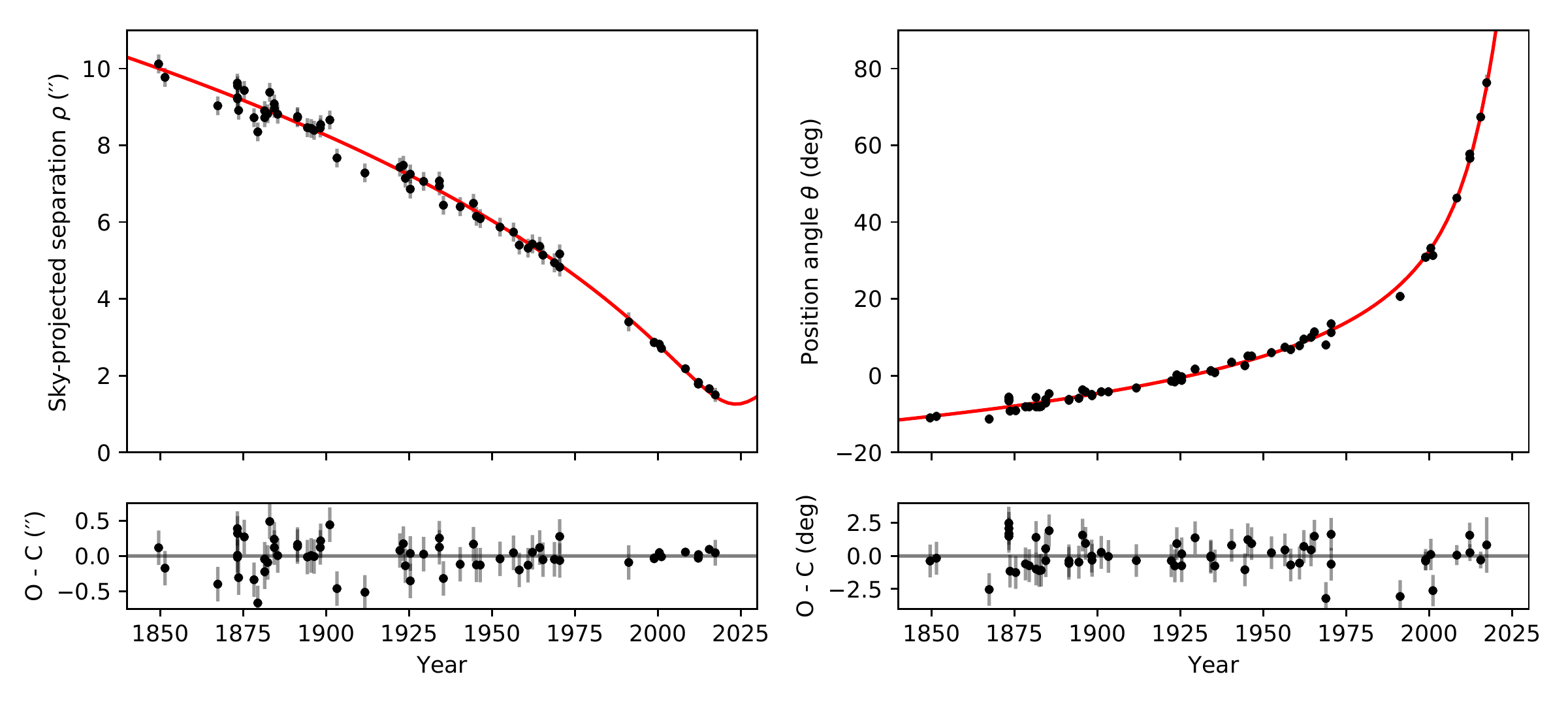}
    \caption{Sky-projected separation (\textit{left}) and position angle (\textit{right}) of \tb\ over the past 168 years. The median model from our MCMC analysis is plotted in red. From the rapidly shrinking separation and accelerating position angle, it is clear that \tb\ B is approaching periastron. Only a small arc of the full orbit is traced by this data.}
    \label{fig:astro_data}
\end{figure*}
    
As a bright ($V =$ \VA) visual double star, \tb\ AB has been recognized as a stellar binary and been measured astrometrically since 1825. Several authors have provided orbits of \tb\ B based on this astrometric data. The earliest orbital solutions were provided by \citet{1994AJ....107..306H} and \citet{1996BABel.153...57P} who found very different solutions of  $P = 2000$\,\textrm{yr}, $e = 0.91$ and $P = 389$\,\textrm{yr}, $e = 0.42$, respectively. Recently, orbital solutions by \citet{2011AJ....142..175R} and \citet{2014AJ....147...65D} find \tb\ B in a highly eccentric ($e=0.76$--$0.84$), $\sim \!\! 1000$\,year orbit. These solutions predict a periastron passage between 2025--2035. \citet{2015A&A...578A..64B} determined an eccentricity of $e = 0.7 \pm 0.2$ by fitting radial velocities while fixing the period and periastron passage using the astrometric values from \citet{2014AJ....147...65D}. 
The primary component \tb\ A (\specA) has been studied extensively for signs of star-planet interactions due to its extremely rapid magnetic cycle of 240 days and likely tidally locked rotation \citep{2005ApJ...622.1075S, 2008ApJ...676..628S, 2008A&A...482..691W, 2009A&A...505..339L, 2012A&A...544A..23L, 2012MNRAS.423.3285V, 2014A&A...565L...1P, 2015A&A...578A..64B, 2018MNRAS.479.5266J}. Measurements of the magnetic field and $S$-index have revealed differential rotation on the order of $\sim \mkern-7mu 20\%$. \citet{2016MNRAS.459.4325M} measured the stellar rotation period using spectropolarimetric observations from 2011--2015 and reports an equatorial angular velocity in the range $\Omega_{\mathrm{eq}} = 1.95$--2.05\,rad/d, corresponding to rotation periods in the range $P_* = 3.06$--$3.34$\,days. At a stellar latitude of $40^{\circ}$, the stellar rotation period corresponds to the 3.3 days orbital period of \tb\ Ab, which suggests that tidal forces have forced the thin convective envelope of \tb\ A into corotation \citep{2009MNRAS.398.1383F}. \citet{2015A&A...578A..64B} detected solar-like oscillations in \tb\ A and confirmed the strong differential rotation. We adopt the stellar parameters of \tb\ A by \citet{2005ApJS..159..141V}. The adopted stellar parameters of \tb\ A are listed in Table \ref{table:stellarparams}.

\tb\ Ab was discovered by \citet{1997ApJ...474L.115B} via the radial velocity method as part of the Lick planet search effort using the Hamilton spectrograph at Lick observatory. \tb\ Ab is a non-transiting hot Jupiter in a 3.3 day orbit. As one of the earliest known and brightest exoplanetary systems, \tb\ Ab is one of the most well-studied exoplanets. The presence of CO \citep{2012Natur.486..502B, 2012ApJ...753L..25R} and water \citep{2014ApJ...783L..29L} has been measured in the absorption spectrum of \tb\ Ab using near-IR spectroscopy. These measurements have enabled the determination of the orbital inclination angle of \tb\ Ab. \citet{2012Natur.486..502B} and \citet{2012ApJ...753L..25R} determined an orbital inclination of $i_{\rm Ab} =$ \inclAb\ and $47^{+7}_{-6}$\,deg, respectively, by tracing the radial velocity shift of planetary CO absorption lines in the near-infrared spectrum of \tb\ Ab. \citet{2014ApJ...783L..29L} determined an orbital inclination of $i_{\rm Ab} = 45^{+3}_{-4}$\,deg by measuring the radial velocity shift of water vapor in \tb\ Ab's atmosphere. For this study, we adopt the orbital inclination by \citet{2012Natur.486..502B}. The knowledge of the orbital inclination has enabled the determination of the mass of \tb\ Ab at $M_{\rm Ab} = 5.95\pm 0.28 M_{\rm Jup}$ \citep{2012Natur.486..502B}. \tb\ Ab has been the target of several searches for reflected light with a current $3\sigma$ upper-limit of the planet-to-star contrast of $1.5 \cdot 10^{-5}$, corresponding to an optical albedo of 0.12 \citep{2018A&A...610A..47H}. 

\section{Observational Data} \label{sec:data}
We collect 192 years of astrometric data from the Washington Double Star Catalog \citep[WDS,][]{2001AJ....122.3466M}. The WDS contains 63 observations of \tb\ spanning the period 1825--2017. The three earliest measures, observed by J.F.W. Herschel between 1825--1831 \citep{1826MmRAS...2..459H, 1833MmRAS...6....1H, 1871MmRAS..38....1H} are outliers and not used in the analysis. The second \textit{Gaia} data release contains the position of \tb\ A and B \citep{2016A&A...595A...1G, 2018A&A...616A...1G}. From the \textit{Gaia} DR2 positions, we compute a separation and position angle of $(\rho, \theta) = (1.662\pm 0.002\,\arcsec, 67.36 \pm 0.07$\,deg) at epoch 2015.5. This leaves us with an astrometric baseline of 168 years including 61 observations from 1849--2017. We show the astrometric data (along with a model, see Sec. \ref{sec:results}) in Figure \ref{fig:astro_data}. The \tb\ system has shown significant change in the past 200 years. Astrometric data from the past two decades show the sky-projected separation approaching a minimum and a rapidly changing position angle, indicating that \tb\ B is close to periastron passage. 

We use radial velocities from the Lick Planet Search program \citep{2014ApJS..210....5F}. The survey was conducted from 1987 to 2011 using the Hamilton spectrograph at the Lick Observatory. \tb\ was continuously observed for the duration of the survey. We exclude data prior to 1995, which is recorded with a separate set-up and feature significantly more scatter than later data. The remaining Lick dataset contains 147 RV measurements taken from 1995 to 2011 with three instrumental setups. We furthermore include publicly available data from the GAPS programme using the HARPS-N spectrograph at the TNG. Under this programme, \tb\ was observed on 12 nights between April 13 and May 14 2013 and on 11 nights between April 3 and July 22 2014 \citep{2015A&A...578A..64B}. The GAPS data were obtained as part of an asteroseismic investigation and is therefore of high cadence with up to 30 spectra a night. We bin the GAPS data and use the median of each night. Finally, we observed \tb\ using HARPS-N on two nights, April 9 and July 4 2017. In total, we use 172 RV measurements spanning 22 years.

\section{Analysis} \label{sec:analysis}
\subsection{Orbital Model} \label{sec:Model}
\begin{table}
    \caption[]{Spectral type, magnitude, coordinates, parallax, proper motion and stellar parameters of \tb\ A.}
    \label{table:stellarparams}
    \centering          
    \begin{tabular}{l c c}
        \hline\hline
        \noalign{\smallskip}
        Parameter (unit) & Value & Reference \\
        \noalign{\smallskip}
        \hline
        \noalign{\smallskip}
        Spectral Type & \specA & 1 \\
        $V$ & \VA & 1   \\
        RA (J2000) & \RAA & 2 \\
        DEC (J2000) & \DECA & 2 \\
        $\pi$ (mas) & \parallax & 2 \\
        $\mu_{\alpha}$ \textit{Gaia} (mas\,yr$^{-1}$) & $-467.9\pm 0.7$ & 2 \\
        $\mu_{\alpha}$ HIP (mas\,yr$^{-1}$) &  $-479.5\pm 0.2$ & 3 \\
        $\mu_{\delta}$ \textit{Gaia} (mas\,yr$^{-1}$) &  $64.7\pm 0.5$ & 2 \\
        $\mu_{\delta}$ HIP (mas\,yr$^{-1}$) &  $53.5\pm 0.1$ & 3 \\
        $T_{\mathrm{eff}}$ (K) & \TeffA  & 4     \\
        $\log g$ (cgs) & \loggA  & 4     \\
        ${\mathrm{[M/H]}}$ (dex) & \FeHA  & 4     \\
        $v \sin i$ \textrm{(k\textrm{m\,s}$^{-1}$)} & \vsiniA  & 4     \\
        $M_*$ $(M_\mathrm{_{\odot}})$ & \MA  & 4     \\
        $R_*$ $(R_\mathrm{_{\odot}})$ & \RA  & 4     \\
        $\text{Age (Gyr)}$ & \ageA  & 4     \\
        $P_{\mathrm{rot}}$ (d) & \ProtA  & 5     \\
        \noalign{\smallskip}
        \hline
    \end{tabular}
    \tablebib{(1)~SIMBAD \citep{2000A&AS..143....9W}; (2)~ \textit{Gaia} DR2 \citep{2016A&A...595A...1G, 2018A&A...616A...1G};
    (3)~Hipparcos, the New Reduction \citep{2007A&A...474..653V}; (4)~\citet{2005ApJS..159..141V}; (5)~Covering literature values, e.g. \citet{2016MNRAS.459.4325M}}
\end{table}

We construct a Keplerian model to simultaneously fit the astrometric and radial velocity data. To solve the three-dimensional orbit, we use the equations of Keplerian motion. The orbit's three-dimensional coordinates may be expressed as \citep[see e.g.][]{2010exop.book...15M}
\begin{align}
    x &= r\left(\cos\Omega \cos (\omega + \nu) - \sin \Omega  \sin (\omega + \nu) \cos i \right), \\
    y &= r\left(\sin\Omega \cos (\omega + \nu) + \cos \Omega  \sin (\omega + \nu) \cos i \right), \\
    z &= r \sin(\omega + \nu)\sin i,
\end{align}
where $r = a\left(1 - e\cos E\right)$, $a$ is semi-major axis, $\Omega$ longitude of the ascending node, $\omega$ argument of periastron and $i$ orbital inclination. $(x, y)$ are the sky-projected coordinates while $z$ is the line-of-sight coordinate. The sky-projected coordinates ($x$, $y$) are converted into the astrometric observables (projected separation $\rho$ and position angle $\theta$) via
\begin{align}
    \rho &=  \sqrt{x^2 + y^2}, \\
    \theta &= \arcsin \left(\frac{y}{\rho}\right).
\end{align}
The astrometric data measure the apparent size of the orbit on the sky. To compute the physical size of the orbit, we scale the orbit by the stellar distance $d = 1/\pi$, using the parallax $\pi$. Radial velocity data help to further constrain the orbital parameters. The line-of-sight velocity is computed as
\begin{align}
    v_z &= K\left( \cos(\omega + \nu) + e\cos \omega \right) + \gamma_i,
\end{align}
where $\gamma_i$ is a velocity offset and $K$ is the radial velocity semi-amplitude, which is a function of the masses $M_1$ and $M_2$, the inclination $i$, the semi-major axis $a$ and eccentricity $e$. In summary, to fit radial velocities and astrometric observables of an orbit, our model requires the parameters $(a, T_0, e, \omega, \Omega, i, M_1, M_2, d, \gamma$).

\subsection{Comparing Model to Data}
The astrometric measurements in the WDS are sourced from different observers using different instrumental setups over a period of more than 150 years. The measurements prior to 1998 are almost exclusively made with micrometers on photographic plates. Only data collected after 1998 have formal uncertainties listed\footnote{The most recent measurement \citep[by][]{2018JDSO...14...78S} is listed without uncertainties. For this measurement we adopt an uncertainty of $0.18\,\arcsec$ in separation (corresponding to half the pixel scale) and 2\,deg in position angle.}. To ensure a proper weighting of the data we fit error terms ${\sigma}_{\rho, \rm pre\ 1998}$ and ${\sigma}_{\theta, \rm pre\ 1998}$ to astrometric data taken prior to 1998. The formal uncertainties on the post-1998 astrometric data ranges from 0.002--0.04$\,\arcsec$ in separation and 0.07--1.0\,deg in position angle. To account for potential systematic uncertainties (due to instrumental and atmospheric effects) we fit jitter error terms ${\sigma}_{\rho, \rm post\ 1998}$ and ${\sigma}_{\theta, \rm post\ 1998}$ to this data.

Similarly, the radial velocities are obtained using different instrumental set-ups. The Lick RVs have a mean uncertainty of 8\,\textrm{m\,s}$^{-1}$, while the HARPS-N data have a mean uncertainty of 1\,\textrm{m\,s}$^{-1}$. Due to intrinsic instrumental effects, atmospheric conditions and stellar variability, these uncertainties do not reflect the true error of the measurements. Using the high-cadence GAPS data, \citet{2015A&A...578A..64B} found peak-to-peak variations of up to $15$\,\textrm{m\,s}$^{-1}$ with typical variability less than $10$\,\textrm{m\,s}$^{-1}$. To account for this additional noise, we include a jitter term $\sigma_{\mathrm{RV,}\ i}$ for each instrumental radial velocity set-up.  In order to combine the four RV datasets, we include four RV offsets $\gamma_i$.

The orbits of the stellar companion \tb\ B and the hot Jupiter \tb\ Ab are fitted simultaneous. The astrometric reflex motion of \tb\ A due to the planet \tb\ Ab is of the order of $\sim \! \! 10\,\mu$as, roughly three orders of magnitude smaller than the estimated astrometric uncertainties. We therefore neglect the astrometric signal of the planet. \citet{1997ApJ...474L.115B} reported a marginal eccentricity of \tb\ Ab at the $1\sigma$ level.  \citet{2012Natur.486..502B} redetermined the orbit and found that when excluding noisy Lick data prior to 1995, an eccentric solution is no longer preferred. \citet{2015A&A...578A..64B}, also excluding early Lick data, determined a small eccentricity of 0.011 at the $2\sigma$ level with an argument of periastron near 90 degrees. However, eccentricities derived from RVs may be biased towards larger values \citep{2011MNRAS.410.1895Z}. In particular, the combination of a small eccentricity and an argument of periastron close to 90 degrees is indicative of a potential observational bias \citep{2012ApJ...744..189A}. We therefore adopt a circular orbit of \tb\ Ab. By neglecting the astrometric contribution and adopting a circular orbit, we reduce the dimensionality by four parameters. For \tb\ B we include the full astrometric and radial velocity signal. The final model contains 24 parameters (listed in Table \ref{table:solution}): six parameters describing the orbit of \tb\ B, three parameters describing \tb\ Ab's orbit, four RV offsets, four RV jitter terms, four astrometric error terms, the stellar masses of \tb\ A and B and finally the distance. We adopt the joint log-likelihood function
\begin{align*}
\ln \mathcal{L} &= \ln \mathcal{L}_{\rm RV} + \ln \mathcal{L}_{\rm \theta} + \ln \mathcal{L}_{\rm \rho} + \ln \mathcal{L}_{\rm priors}, \text{ where} \\
\ln \mathcal{L}_{\rm x} &=  -\frac{1}{2} \sum_i \left( \frac{\left( \mathrm{x}_{\mathrm{obs}, i} - \mathrm{x}_{\mathrm{model}, i} \right)^2} {\sigma_{\mathrm{obs}, i}^2 + \sigma_{\rm jit}^2} + \ln 2 \pi \left(\sigma_{\mathrm{obs}, i}^2 + \sigma_{\rm jit}^2 \right) \right) 
\end{align*}
and $\ln \mathcal{L}_{\rm priors}$ is the combined log-likelihood of the priors described below. We impose Gaussian priors on the primary stellar mass $M_{\rm A}$ and distance $d$ (using the values in Table \ref{table:stellarparams}). We impose a $\sin i$ prior on the inclination, appropriate for an isotropic orientation. We use a log-normal prior with a mean of $\log_{10} (P\rm / days) = 5.03$ and a standard deviation of $\sigma_{\log_{10} P/\rm days} = 2.28$ on the period of the binary, consistent with the period distribution of binaries determined by \citet{2010ApJS..190....1R}. For the remaining parameters we assume uniform priors. We sample the posterior distribution using the affine invariant MCMC ensemble sampler $\texttt{emcee}$ \citep{2013PASP..125..306F}.

We run the MCMC algorithm for \nsteps\ iterations using \nwalkers\ parallel chains for a total of \totalsamples\ samples. We reject the initial \nburn\ steps of each chain and thin by a factor of \nthin.

\begin{figure}
    \centering
    \includegraphics[width=\columnwidth]{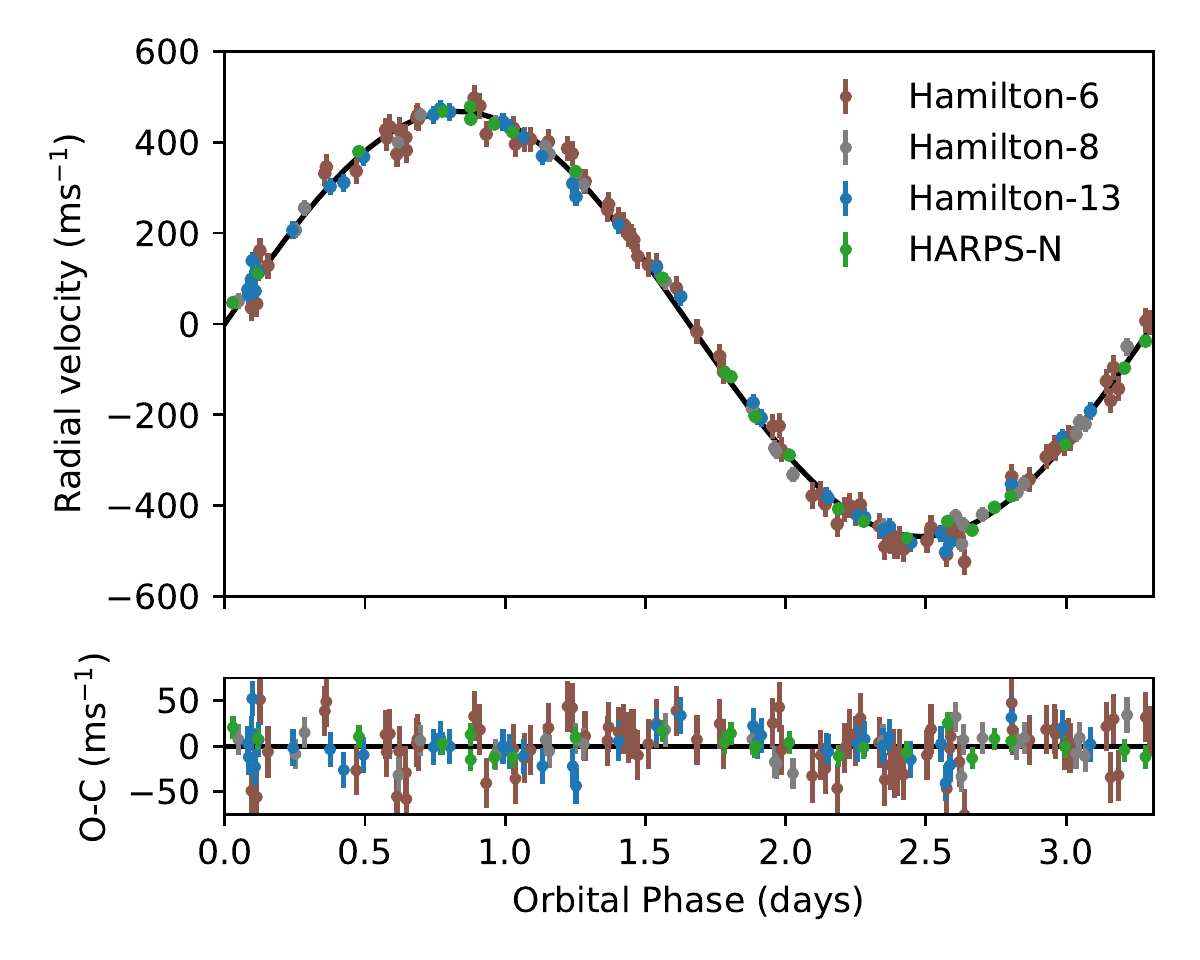}
    \caption{Phase-folded radial velocity curve of \tb\ A without the contribution from \tb\ B. The median model is plotted in black.}
    \label{fig:planet_RV}
\end{figure}

\begin{table}
    \caption{Parameters of the \tb\ system.}            
    \label{table:solution}
    \centering          
    \begin{tabular}{lc}
        \hline\hline
        \noalign{\smallskip}
        Parameter & Value \\
        \noalign{\smallskip}
        \hline
        \noalign{\smallskip}
        \textbf{Model parameters of \tb\ AB} \smallvspace \\ 
        $\log (a/\rm \arcsec)$ & \logaarc \smallvspace \\ 
        $t_0$ (MJD) & \tperiMJD \smallvspace \\ 
		$\sqrt{e} \cos \omega$ & \ecosw \smallvspace \\      
		$\sqrt{e} \sin \omega$ & \esinw \smallvspace \\      
        $\Omega$ (deg) & \bigomega \smallvspace \\ 
        $\cos i$ & \cosi \smallvspace \\ 
        $M_{\mathrm{A}}$ $(M_{\odot})$ & \MA \smallvspace \\ 
        $M_{\mathrm{B}}$ $(M_{\odot})$ & \MB \smallvspace \\
        $d$ (pc) & \distance \smallvspace \\
        \hline \noalign{\smallskip} \smallvspace
        \textbf{Derived parameters of \tb\ AB} \smallvspace \\
        $R_{\rm B}$\tablefootmark{$\dagger$} ($R_{\odot}$) & \RB \smallvspace \\ 
        $T_{\rm eff, B}$\tablefootmark{$\dagger$} (K) & \TeffB \smallvspace \\ 
	    $M_{\rm A} + M_{\rm B}$ $(M_{\odot})$ & \athreePtwo\ \smallvspace \\ 
        $t_0$ (AD) & \tperiAD \smallvspace \\ 
        $P$ (\textrm{yr}) & \Pyr \smallvspace \\
        $a$ (\textrm{au}) & \aAU \smallvspace \\ 
        $a_{\rm peri}$ (\textrm{au}) & \aAUperi \smallvspace \\ 
        $e$ & \ecc \smallvspace \\ 
        $\omega$ (deg) & \smallomega \smallvspace \\ 
        $i$ (deg) & \incl \smallvspace \\ 
        \hline \noalign{\smallskip} \smallvspace
        \textbf{Model parameters of \tb\ Ab} \smallvspace \\ 
        $K_{\rm Ab}$ (\textrm{m\,s}$^{-1}$) & \Kp \smallvspace \\ 
        $P_{\rm Ab}$ (days)& \Pp \smallvspace \\
        $t_{0, \rm Ab}$ (MJD) & \tp \smallvspace \\ 
        \hline \noalign{\smallskip} \smallvspace
        \textbf{Additional model parameters} \smallvspace \\
        $\gamma_{\rm Ham13}$ (\textrm{m\,s}$^{-1}$) & \gaa \smallvspace \\ 
        $\gamma_{\rm Ham6} - \gamma_{\rm Ham13}$ (\textrm{m\,s}$^{-1}$) & \gab \smallvspace \\ 
        $\gamma_{\rm Ham8} - \gamma_{\rm Ham13}$ (\textrm{m\,s}$^{-1}$) & \gac \smallvspace \\ 
        $\gamma_{\rm HARPS-N} - \gamma_{\rm Ham13}$ (\textrm{m\,s}$^{-1}$) & \gad \smallvspace \\ 
        $\sigma_{\rm RV,\ Ham13}$ (\textrm{m\,s}$^{-1}$) & \jrva \smallvspace \\ 
        $\sigma_{\rm RV,\ Ham6}$ (\textrm{m\,s}$^{-1}$) & \jrvb \smallvspace \\ 
        $\sigma_{\rm RV,\ Ham8}$ (\textrm{m\,s}$^{-1}$) & \jrvc \smallvspace \\ 
        $\sigma_{\rm RV,\ HARPS-N}$ (\textrm{m\,s}$^{-1}$) & \jrvd \smallvspace \\ 
        $\sigma_{\rho,\ \rm pre\ 1998}$ (\arcsec) & \jrhoold \smallvspace \\ 
        $\sigma_{\theta,\ \rm pre\ 1998}$ (deg) & \jthetaold \smallvspace \\ 
        $\sigma_{\rho,\ \rm post\ 1998}$ (\arcsec) & \jrhonew \smallvspace \\ 
        $\sigma_{\theta,\ \rm post\ 1998}$ (deg) & \jthetanew \smallvspace \\ 
        \noalign{\smallskip}
        \hline
    \end{tabular}
    \tablefoot{\tablefoottext{$\dagger$}{From isochrone fitting}}
\end{table}

\section{Results} \label{sec:results}
\begin{figure*}
    \centering
    \includegraphics[width=\textwidth]{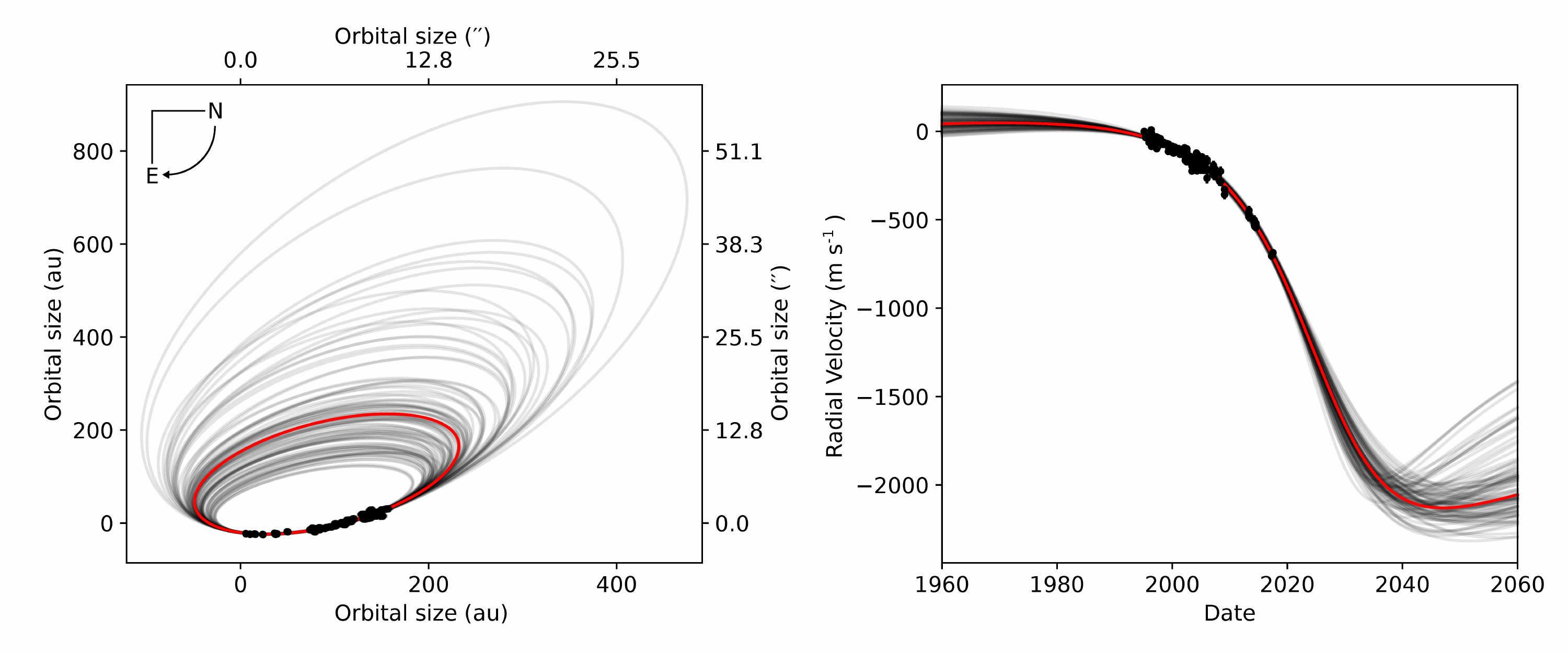}
    \caption{\textit{Left panel}: Sky-projected relative orbit of \tb\ AB in apparent size (\arcsec) and absolute size (\textrm{au}). See Figure \ref{fig:astro_data} for a closer view of the astrometric data and model residuals. \textit{Right panel:} Radial velocity curve of \tb\ A with planetary RV signal subtracted. See Figure \ref{fig:planet_RV} for a closer view of the radial velocity curve and model residuals.}
    \label{fig:orb_samples}
\end{figure*}

The results of the analysis are listed in Table \ref{table:solution}. We report median values with $1\sigma$ ($16\%$ and $84\%$ quantile) uncertainties. The median model is plotted in Figure \ref{fig:astro_data} along with the astrometric data and residuals, and in Figure \ref{fig:planet_RV} with radial velocity data and residuals. In Figure \ref{fig:orb_samples}, the sky-projected orbit and radial velocity signal are plotted with the median and 100 orbits drawn from the posterior. \tb\ B is in a highly eccentric ($e= $\ecc) orbit. As evident from the rapidly shrinking separation and the significant change in the RV amplitude in the past 25 years, \tb\ AB is rapidly approaching periastron. We predict periastron in year \tperi. The period ($P=$ \Pyr\,yr) and semi-major axis ($a=$ \aAU\,\textrm{au}) are relatively poorly determined. This is a general feature of orbits with low orbital coverage, since it is possible to draw an almost arbitrarily large ellipse through the data. However, the data close to periastron passage enable a precise determination of the periastron distance of \aAUperi\,\textrm{au}. The combination of astrometry and RVs constrains the three-dimensional orientation of the orbit, allowing constraints on $\Omega = $ \bigomega\,deg, $\omega = $ \smallomega\,deg and $i = $ \incl\,deg. While the period and semi-major axis are relatively poorly constrained, the relation $(a/\mathrm{\textrm{au}})^3/(P/\mathrm{yr})^2=$ \athreePtwo, equivalent to the mass sum in solar units via Kepler's third law, is much more precise. This was noted by \citet{1967ARA&A...5..105E} who discovered that is it possible to derive reliable mass sums even for binaries with low orbital coverage. Combining the mass sum with the mass of \tb\ A ($M_{\rm A} = $ \MA $M_{\odot}$) we get a precise mass of \tb\ B $M_{\rm B} = $ \MB\ $M_{\odot}$. Assuming that the two stellar companions have the same age and metallicity, we determine the radius of \tb\ B by fitting the age, metallicity and mass to a grid of BaSTI isochrones \citep{2004ApJ...612..168P} using the BAyesian STellar Algorithm, \texttt{BASTA} \citep{2015MNRAS.452.2127S}. We find $T_{\text{eff, B}} = $ \TeffB\,K and $R_{\text{B}} = $ \RB\ $R_{\odot}$. The effective temperature is in good agreement with the temperature $T_{\text{eff, B}} = 3425 \pm 100$\,K determined by \citet{2016A&A...586A.100L} using high-resolution infrared spectroscopy. 

As a consistency check of our solution, we check that the change in proper motion (\textit{i.e.} sky-projected velocity) of \tb\ A between Hipparcos and \textit{Gaia} DR2 is consistent with the change expected from our solution. We predict a change in the proper motion of \tb\ A due to its astrometric reflex motion caused by \tb\ B. We note that the proper motions reported by Hipparcos and \textit{Gaia} DR2 are derived by assuming a constant proper motion for the duration of their observations. This assumption does not hold for \tb\ A, since the magnitude of its proper motion changes by $\sim \! \! 1.6$\,mas\,yr$^{-1}$ during \textit{Gaia} DR2 observations and  $\sim \! \! 1.2$\, mas\,yr$^{-1}$ during Hipparcos observations. The Hipparcos and \textit{Gaia} DR2 measurements are therefore likely not as accurate as the reported formal uncertainties suggest. To compare with our model, we compute the mean proper motion of \tb\ A over the observing periods of Hipparcos (1989.8--1993.2) and \textit{Gaia} DR2 (2014.6--2016.4). From our analysis, we find that the magnitude and angle of the proper motion difference $\Delta \mu$ of \tb\ A are $\Delta \mu = \sqrt{\Delta \mu_{\alpha}^2 + \Delta \mu_{\delta}^2} = 14.2 \pm 0.8$\, mas\,yr$^{-1}$ at an angle $\theta = \arctan(\Delta\mu_{\alpha}/\Delta\mu_{\delta}) = 47.6\pm 2.4$\,deg. The magnitude and angle of the proper motion difference between the Hipparcos and \textit{Gaia} DR2 catalogs are $\Delta \mu = 16.1\pm 0.9$\,mas\,yr$^{-1}$ at an angle $\theta = 46.0\pm 2.3$\,deg, in good agreement with our solution.
\begin{figure}
  \centering
  \includegraphics[width=\columnwidth]{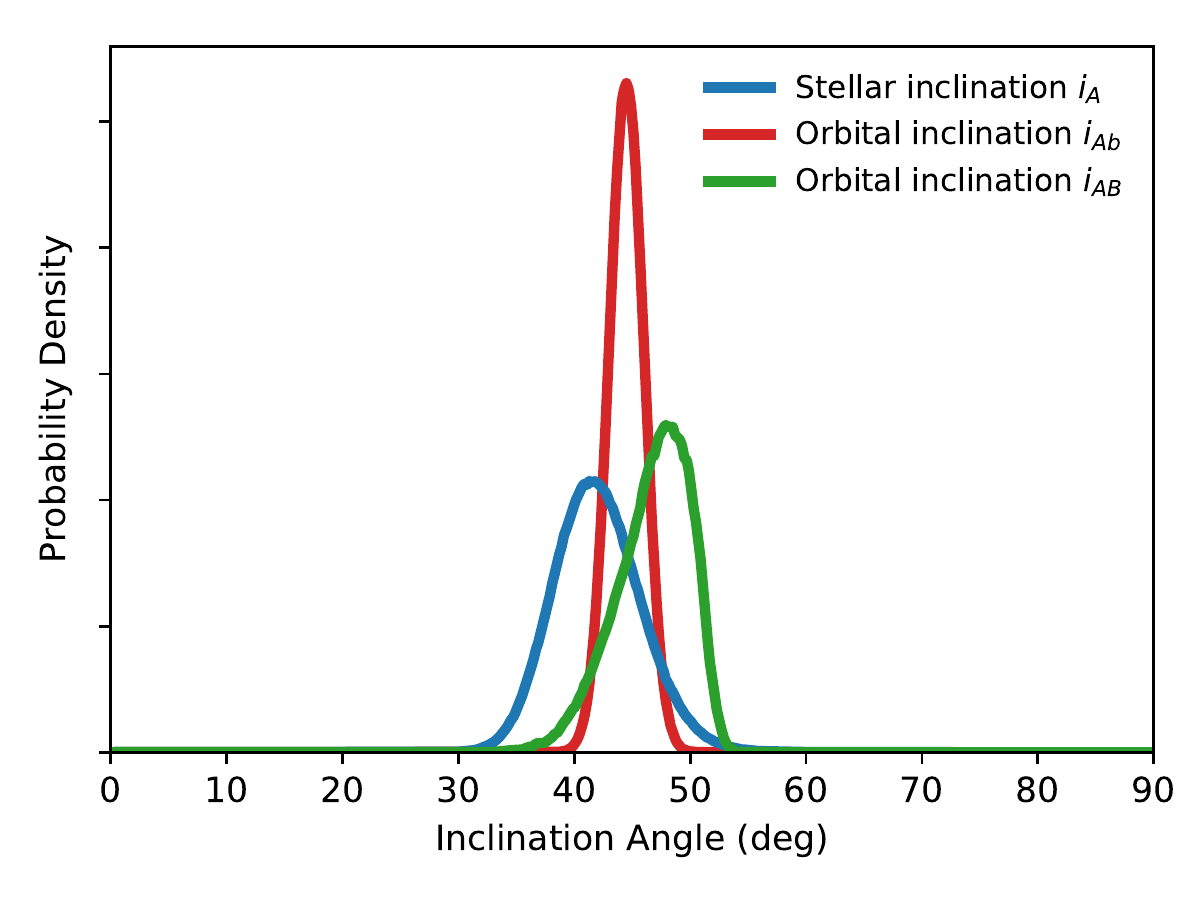}
  \caption{Probability density of the inclination of the stellar spin-axis of \tb\ A $i_A$ (derived from $v\sin i$, $R_*$ and $P_{\rm rot}$, see Table \ref{table:stellarparams}), the orbital inclination angle of \tb\ Ab $i_{Ab}$ \citep[determined via the planetary RV signal by][]{2012Natur.486..502B} and the orbital inclination angle of \tb\ B $i_{AB}$ (determined from our MCMC analysis). The similarity of the three inclination angles suggests a coplanar, aligned system.}
  \label{fig:incl_dist}
\end{figure}

\section{Discussion} \label{sec:discussion}
\subsection{Mutual inclinations in the \tb\ system}
The orbital inclination of the stellar companion \tb\ B is important for understanding the dynamical evolution of the \tb\ system. If the stellar companion triggered the migration of the hot Jupiter \tb\ Ab via the Kozai-Lidov mechanism, \tb\ Ab and \tb\ B are expected to orbit on mutually inclined orbits. In classic Kozai-Lidov theory, the minimum mutual inclination required to excite an eccentricity $e_{\rm max}$ is \citep{1997Natur.386..254H}
\begin{align}
    i_{\rm mutual} = \arccos \left( \sqrt{\frac{3}{5}(1-e_{\rm max}^2)}\right) \geq 39.2 \rm\,deg.
\end{align}
From the joint astrometric and RV analysis we determine an orbital inclination of \tb\ B of $i_{\rm B} = $ \incl\,deg. This value is consistent with the orbital inclination of \tb\ Ab ($i_{\rm Ab} =$ \inclAb), suggesting a coplanar configuration with at most a small mutual inclination $i_{\rm B - Ab} = 2.6_{-3.9}^{+3.1}$\,deg. The mutual inclination is too small to have triggered Kozai-Lidov oscillations. To investigate the spin-orbit alignment, we determine the inclination angle of the stellar spin-axis of \tb\ A by combining the stellar radius, rotation period and projected rotational velocity (as given in Table \ref{table:stellarparams})
\begin{align}
    i_{\rm A} &= \arcsin \left(\frac{v \sin i_{A} \cdot P_{\mathrm{rot,}\ A}}{2 \pi R_{*,\ A}}\right) = 42 \pm 4\,\textrm{deg}.
\end{align}
We find that all three inclination angles agree within their uncertainties. The probability density functions of the three inclination angles are plotted in Figure \ref{fig:incl_dist}. The three coinciding inclination angles are consistent with the assumption of a flat system architecture in which the orbits of the companions \tb\ Ab and \tb\ B are both aligned within a few degrees of the stellar spin-axis of \tb\ A. We note that the inclination angles measure only the projected alignment along the line of sight. The longitude of the ascending node of the orbit of \tb\ Ab, as well as the sky-projection of the stellar spin-axis of \tb\ A are unconstrained by the data. Since the sky-projected alignment is unconstrained, it is possible that the system is non-coplanar and misaligned in three-dimensional space. It is however unlikely that all three inclination angles would agree by coincidence. The assumption of coplanarity and alignment is the simplest (but not the only) interpretation of the data. In the following section we therefore assume that the system is coplanar and aligned and discuss the implications of this assumption.

It is unlikely that tidal forces have realigned the orbits or spin-axis in the system. If \tb\ Ab has been tidally realigned with the stellar spin-axis, we would not expect that the stellar spin-axis is aligned with the outer companion. Additionally, due to its shallow convective envelope, the tidal forces of \tb\ A are expected to be too weak to realign a misaligned orbit. Using the calibrated timescale for changing the spin of a star from \citet{2012ApJ...757....6H} and assuming a stellar gyration radius $k_0^2 = 0.1 R_*^2$, a stellar dissipation coefficient $\bar{\sigma}_* = 10^{-8}$ (appropriate for a 1.35$M_{\odot}$ star) and a planet radius of 1.15$R_{\rm Jup}$, we find a time scale of $T_{\mathrm{spin}} \sim 10^{11}$\,yr, much longer than the $\sim \! \! 1$\,Gyr age of the \tb\ system. We therefore conclude that both \tb\ Ab and \tb\ B formed and migrated within a primordially aligned disc. Using Eq. 1 in \citet{1999AJ....117..621H}\footnote{Extrapolating their results slightly from $e = 0.8$ to $e = 0.87$}, we find that planetary orbits in the \tb\ AB binary are stable out to $2\%$ of the binary semi-major axis, a condition easily satisfied by \tb\ Ab which orbits at a distance less than $0.05\%$ of the binary semi-major axis.

\subsection{Truncation of the protoplanetary disc}
The formation of circumstellar (S-type) planets around close or highly eccentric binaries presents a significant challenge for current formation theories. In a close binary, the protoplanetary disc around the primary star will be tidally truncated by the companion at a fraction of the periastron distance \citep{1994ApJ...421..651A}. The location of the truncation radius depends on the separation, mass ratio, eccentricity and inclination of the companion.

Due to the high eccentricity of \tb\ B, the periastron approach may be approximated as a parabolic stellar fly-by. \citet{1993MNRAS.261..190C} studied the influence of a single coplanar fly-by and found that the disc is typically stripped of material down to $\approx0.5a_{\rm peri}$. \citet{2014A&A...565A.130B}, also studying stellar fly-bys, found a simple relation $a_{\rm trunc} = 0.28 \mu^{-0.32} a_{\rm peri} \approx 0.4 a_{\rm peri}$ between the truncated disc size $a_{\rm trunc}$ and the mass-ratio $\mu$ and periastron distance $a_{\rm peri}$ of the fly-by. \citet{2016A&A...594A..53B} showed that inclined encounters truncate the protoplanetary disc slightly less efficiently than coplanar encounters, with an approximately linearly decreasing effect on final disc size with increasing star-disc inclination. It is clear that even a single encounter truncates the disc at a fraction of the periastron distance. 

\citet{2005MNRAS.359..521P} studied the discs of stars in binaries by simulating the effect of periodic gravitational perturbations on test particle orbits. From their numerical simulations they derive an equation estimating the outer truncation radius of circumstellar discs in binaries
\begin{align}
    a_{\rm trunc} = \frac{0.359 q_1^{2/3} (1-e)^{0.20} q^{0.07}}{0.6q_1^{2/3} + \ln\left(1 + q_1^{1/3}\right)} a_{\rm peri},
\end{align}
where $q = m_2/(m_1 + m_2)$ and $q_1 = (1 - q)/q$. For \tb\ A, we find a truncation radius of $a_{\rm trunc} \approx 0.2 a_{\rm peri}\approx 6$\,\textrm{au}. This truncation radius is a factor of two smaller than the expected truncation radius caused by a single stellar encounter.

\subsection{Planet formation and migration in a truncated disc}
Not only is the protoplanetary disc truncated, the shape and velocity dispersion of the disc is also affected. The increased velocity dispersion in a perturbed disc may prohibit planetesimal growth. \citet{2001Icar..153..416K} studied the velocity dispersion of planetesimals in a circular disc after a stellar encounter and find that planet formation is inhibited in the outer disc due to increased relative impact velocities. They derive the outer boundary of the planet-forming radius $a_{\rm planet}$ as the distance below which the velocity dispersion is low enough for planet formation. They find the relation
\begin{align}
    a_{\rm planet} \approx 40 \left(\frac{\mu + \mu^{-2}}{2}\right)^{1/4}\left(\frac{a_{\rm peri}}{150\,\textrm{au}}\right)^{5/4}\,\textrm{au} \approx 7\,\textrm{au},
\end{align}
where $\mu = m_1/m_2$. The estimated outer radius of the planet-forming region is comparable to the truncation radius, suggesting that planet formation is possible in the truncated disc. The eccentricity of the disc is another a problem for efficient planet formation. In an eccentric disc, relative impact velocities are increased, impeding (or potentially preventing) grain accumulation \citep{2006Icar..183..193T}. However, simulations show that disc self-gravitation acts to reduce relative impact velocities and may enable planetesimal growth in binaries with massive discs \citep{2009A&A...508.1493M, 2013ApJ...765L...8R}. \citet{2009A&A...508.1493M} and \citet{2011A&A...528A..93R} found that circumstellar disc eccentricity decreases for increasing binary eccentricity. The conditions for planet formation may therefore be more favourable in eccentric binaries than compact circular binaries.

\citet{2008A&A...486..617K} studied the migration of protocores in the circumstellar disc of $\gamma$ Cephei A. $\gamma$ Cephei is an eccentric, planet-hosting binary ($a=18.5$\,\textrm{au}, $M_{\rm A} = 1.59M_{\odot}$, $M_{\rm B} = 0.38M_{\odot}$, $e=0.36$ binary hosting a $a = 2.1$\,\textrm{au}, $m\sin i = 1.7M_{\rm Jup}$ planet) in a coplanar configuration \citep{2003ApJ...599.1383H}. Their simulations show that perturbations from $\gamma$ Cephei B periodically generates strong spiral waves propagating towards the center of the circumstellar disc around $\gamma$ Cephei A. They found that gas accretion in such a system is efficient, with a $3M_{\rm Jup}$ disc sufficient to create a $2M_{\rm Jup}$ planet. Embedded protocores initially placed within $a < 2.7$\,\textrm{au} migrate inwards with limited eccentricities, while cores placed at larger distances acquire large eccentricities while slowly migrating inwards. The location of this transition boundary depends on disc mass, viscosity and temperature. \citet{0004-637X-798-2-70} similarly found that is it possible to form giant planets at \textrm{au}-scale orbits in $\sim \! \! 20$\,\textrm{au}-separation eccentric binaries given that the disc is massive and nearly circular.

There is some observational evidence that companion stars excite spiral arms in protoplanetary discs. The protoplanetary disc around HD100453A features a two-armed spiral structure \citep{2015ApJ...813L...2W}. \citet{2016ApJ...816L..12D} and \citet{2018ApJ...854..130W} found that the observed spiral arm structure is most likely generated by the mildly eccentric coplanar stellar companion at a separation of $\sim \! 120$\,\textrm{au}. The stellar companion has truncated the disc of HD100453A at $\sim \! 40$\,\textrm{au}, in good agreement with the numerical simulations discussed earlier. \citet{2016ApJ...827....8N} found that while hot Jupiter host stars have a $\sim \! 3$ times larger companion fraction in the range 50--2000\,\textrm{au} than field stars, the stellar companions are unlikely to cause Kozai-Lidov migration. This suggests that some other mechanism produces hot Jupiters in these binaries.

Spiral density wave-induced disc migration may explain the presence of hot Jupiters in binary stars with small periastron distances. In this picture, a protocore forms at the ice line near the outer edge of the truncated disc and acquires a large eccentricity while efficiently accreting gas until being tidally circularized in a short period orbit. Density wave-induced spiral arms in the disc generated by the stellar companion facilitates both the formation of the protocore (due to increased surface density in the spiral arms)  and trigger inwards disc migration, either eccentric or circular depending on where in the disc the core forms. Core formation is typically most efficient just outside the ice line. Inside the ice line, gases cannot condense into grains needed for efficient core growth\footnote{An exception is in very massive discs where growth can occur even inside the ice line \citep{2004ApJ...604..388I}.}. At large distances core accretion is too slow. The location of the ice line of \tb\ A is estimated as \citep{2005ApJ...626.1045I} 
\begin{align}
    a_{\rm ice} \approx 2.7\left(\frac{M_*}{M_{\odot}}\right)^2\,\textrm{au} \approx 5\,\textrm{au}.
\end{align}
The location of the ice line is comparable to the estimated planet-forming region and the disc truncation radius, indicating that planet formation occurs near the disc edge. The high-density spiral arms may offer favourable conditions for rapid core growth. Core growth by e.g. pebble accretion depends strongly on metallicity and surface density \citep{2014A&A...572A.107L}. The high metallicity of \tb\ A $(\rm{[Fe/H] = 0.25\,dex)}$ enables rapid assembly of the protocore, which is needed in order for the protocore to accrete enough gas before the disc disperses. A scenario in which the massive planet formed near $5\, $\textrm{au} and migrated inwards is consistent with the predictions from pebble accretion \citep{2015A&A...582A.112B}.

Assuming an accretion efficiency similar to $\gamma$ Cephei A, a disc mass of at least $\approx 8M_{\rm Jup}$ is required to form a massive hot Jupiter such as \tb\ Ab (at $6M_{\rm Jup}$). This requires a disc-to-star mass ratio of $M_{\rm d}/M_* \approx 0.5\%$ of the truncated disc.  \citet{2013ApJ...771..129A} report typical values of $M_{\rm d}/M_* \approx 0.2\%- 0.6\%$ for single stars with full discs but with a distribution reaching $M_{\rm d}/M_* \approx 10\%$ at the highest end. If \tb\ A had an initially massive disc of $0.1M_* \approx 140M_{\rm Jup}$, the disc could have lost $95\%$ of its mass while still providing sufficient material for the formation of \tb\ Ab. However, the luminosity of discs around components in binaries with separations $<300$\,\textrm{au} are typically a factor of five reduced compared to single stars \citep{1994ApJ...429L..29J, 1996ApJ...458..312J, 2012ApJ...751..115H}, suggesting that massive discs are uncommon in compact binaries. 

If spiral wave-induced disc migration is a dominant source of hot Jupiters in stellar binaries, we expect that hot Jupiters in binary systems are tracers of the primordial disc alignment. Hot Jupiters born in binaries with aligned discs are therefore expected to be spin-orbit aligned. However, protoplanetary discs in binary systems are not always well-aligned. The protoplantary discs in the young binary HK Tauri are misaligned by $\sim \! \! 60$\,degrees, indicating that one or both discs are misaligned with the binary orbit \citep{2014Natur.511..567J}. The protoplanetary discs in the binary protostar IRS 43 are similarly misaligned with respect to the binary plane \citep{2016ApJ...830L..16B}. An inclined binary companion may cause the primary protoplanetary disc to precess and eventually become misaligned with the stellar spin-axis \citep{2012Natur.491..418B, 2014MNRAS.440.3532L}. \citet{2017ApJ...844..103T} found strong orbital alignment in triple star systems with outer projected separations less than $\sim \!\! 50$\,\textrm{au} and misalignment in systems with outer orbits wider than $1000$\,\textrm{au}. \citet{1994AJ....107..306H} similarly found spin-orbit alignment in binaries with separations less than 30--40\,\textrm{au}. This suggests that hot Jupiters in compact binaries are more likely to be spin-orbit aligned. The Catalog of Exoplanets in Binary Star Systems\footnote{\url{www.univie.ac.at/adg/schwarz/multiple.html}} \citep{2016MNRAS.460.3598S} lists 13 planet-hosting binaries with projected separations less than $50$\,\textrm{au}. Out of these 13 systems\footnote{According to the Orbital Obliquity Catalogue at \url{www.astro.keele.ac.uk/jkt/tepcat/} \citep{2011MNRAS.417.2166S}.}, only WASP-11Ab (with a separation of 42\,\textrm{au}) has a measured projected spin-orbit alignment of $\lambda = 7\pm 5$\,deg \citep{2015A&A...579A.136M}. WASP-11Ab likely formed in a truncated disc similar to $\tau$ Boo Ab. However, we note that due to its low effective temperature ($T_{\rm eff} =  4900 \pm 65$\,K), tidal forces could have realigned the stellar spin-axis of WASP-11A. The Transiting Exoplanet Survey Satellite (TESS) started science operations on July 25, 2018 and will complete a near all-sky survey in two years \citep{2015JATIS...1a4003R}. TESS will increase the number of bright planets amenable for follow-up observations and will likely expand the sample of known planets in compact binaries.

Due to the challenges of forming planets in discs around close binaries, several alternative formation scenarios have been proposed. \citet{2018MNRAS.478.4565G} investigated the conversion of a $P$-to-$S$-type planet by scattering-induced capture. In this scenario, a circumbinary planet in a close ($a=0.5$--3\,\textrm{au}) binary is dynamically scattered, then tidally captured by the primary or secondary star. However, the probability of such an event is highest for small mass-ratios and compact, nearly circular binaries. This scenario is therefore unlikely as an formation path for \tb\ Ab. Another possibility is that the current configuration of the binary is different from the configuration at planet formation. If the binary was less eccentric at the time of planet formation, the protoplanetary disc of \tb\ A would have lost less mass and the formation of massive planets would have been more favourable. \citet{2007A&A...467..347M} studied the orbital evolution of binary stars due to stellar encounters and found that a single stellar encounter may significantly change the orbit of a binary without ejecting planets. However, such an encounter will typically alter the inclination of the binary orbit. The aligned, coplanar configuration of the \tb\ system suggests that it has not been subject to external perturbations or three-body interactions since formation. Furthermore, the probability of a sufficiently close stellar encounter is extremely small for field stars. Another possibility is that the eccentricity of \tb\ B has been raised by star-disc interactions after or during formation of the planet. \citet{1991ApJ...370L..35A} found that star-disc interactions rapidly excite the eccentricity of proto-binaries while in the embedded disk phase. The growth of binary eccentricity occurs before disk dispersal. The eccentricity of \tb\ B is therefore unlikely to have changed significantly after dispersal of the disk.

The combination of astrometry and radial velocities is a powerful method for constraining three-dimensional orbits. When \textit{Gaia} epoch astrometry is released it will be possible to fit the relative motion of \tb\ A and B and similar binaries. This will help to further constrain the orbits of long period binaries. The $Gaia$ mission is expected to discover tens of thousands of planets based on high-precision astrometric measurements  \citep{2014ApJ...797...14P}. By combining $Gaia$ astrometry with radial velocities, it will become possible to obtain three-dimensional solutions of binary star and exoplanet orbits. This will be a valuable tool for breaking the degeneracy between the projected orbital alignment and the true three-dimensional alignment in multi-planet systems and planet-hosting binaries.

\section{Conclusions} \label{sec:conclusions}
By determining the inclination, eccentricity and periastron distance of \tb\ B we find an orbital architecture of the \tb\ system consistent with the assumption of an aligned, coplanar configuration. This configuration suggests that we are observing the primordial configuration of the system and that dynamical interactions such as the Kozai-Lidov mechanism have not played a role in this system. We suggest a likely formation pathway for \tb\ Ab and similar hot Jupiters: The stellar companion (with a periastron distance of $\sim \! \!30$\,\textrm{au}) truncated the protoplanetary disc of \tb\ A near the ice line, reducing the disc to a fraction of its initial size. \tb\ Ab likely formed near the outer edge of this truncated disc and migrated inwards in the disc due to spiral waves generated by the eccentric stellar companion. During migration, \tb\ Ab acquired a large eccentricity that was tidally dampened by \tb\ A, creating the hot Jupiter we observe today. This scenario requires an initially massive protoplanetary disc, rapid assembly of the protocore and highly efficient gas accretion. The existence of planets in compact and eccentric binaries such as \tb\ and $\gamma$ Cephei proves that giant planet formation is possible even in these challenging environments. The discovery and characterisation of the orbital architecture of planet-hosting binaries such as \tb\ are crucial for informing models of planet formation and migration. 
\begin{acknowledgements}
    We thank Davide Gandolfi and Artie Hatzes for the time exchange that made our HARPS-N observations possible. We thank Anders Johansen for discussions that improved this article. We thank the anonymous referee for quick and useful comments that improved the clarity of the manuscript. We acknowledge support from the Danish Council for Independent Research, through a DFF Sapere Aude Starting Grant no. 4181-00487B. Funding for the Stellar Astrophysics Centre is provided by The Danish National Research Foundation (Grant agreement no.: DNRF106). Based on observations made as part of the observing programs OPT17A\_64 and A35TAC\_26 with the Italian Telescopio Nazionale Galileo (TNG) operated on the island of La Palma by the Fundación Galileo Galilei of the INAF (Istituto Nazionale di Astrofisica) at the Spanish Observatorio del Roque de los Muchachos of the Instituto de Astrofisica de Canarias. This project has received funding from the European Union's Horizon 2020 research and innovation programme under grant agreement No 730890. This material reflects only the authors views and the Commission is not liable for any use that may be made of the information contained therein.  This research has made use of the Washington Double Star Catalog maintained at the U.S. Naval Observatory. This work has made use of data from the European Space Agency (ESA) mission
    {\it Gaia} (\url{https://www.cosmos.esa.int/gaia}), processed by the {\it Gaia}
    Data Processing and Analysis Consortium (DPAC,
    \url{https://www.cosmos.esa.int/web/gaia/dpac/consortium}). Funding for the DPAC
    has been provided by national institutions, in particular the institutions
    participating in the {\it Gaia} Multilateral Agreement. This research made use of NumPy \citep{van2011numpy}, SciPy \citep{scipy} and corner.py \citep{corner}. This research made use of Astropy, a community-developed core Python package for Astronomy \citep{2018AJ....156..123T}. This research made use of matplotlib, a Python library for publication quality graphics \citep{Hunter:2007}. This research has made use of the SIMBAD database, operated at CDS, Strasbourg, France \citep{2000A&AS..143....9W}. This research has made use of NASA's Astrophysics Data System. 
\end{acknowledgements}

\bibliography{main.bib}{}
\bibliographystyle{aa}
\end{document}